\documentclass[10pt]{article}
\usepackage{latexsym}
\usepackage{graphicx}
\usepackage{multirow}
\usepackage{amsmath}
\usepackage{ulem}
\usepackage{setspace}
\usepackage{subfigure}
\usepackage{color}
\usepackage{soul}
\begin{document}
\begin{center}\Large{A Bohr-type model of a composite particle using gravity as the attractive force}
\end{center}
\begin{center}
\renewcommand\thefootnote{\fnsymbol{footnote}}
C.G. Vayenas$^{1,2,}$\footnote{E-mail: cgvayenas@upatras.gr; T.Fokas@damtp.cam.ac.uk}, S. Souentie$^1$ \& A. Fokas$^{2,3,^*}$
\end{center}
\begin{center}
\textit{$^1$LCEP, Caratheodory 1, St., University of Patras, Patras GR 26500, Greece\\$^2${Division of Natural Sciences, Academy of Athens, Panepistimiou 28, Ave., GR-10679 Athens, Greece}\\$^{3}${Department of Applied Mathematics and Theoretical Physics, University of Cambridge, Cambridge, CB3 0WA, UK}} 
\end{center}

\begin{abstract}
We formulate a Bohr-type rotating particle model for three light particles of rest mass $m_o$ each, forming a bound rotational state under the influence of their gravitational attraction, in the same way that electrostatic attraction leads to the formation of a bound proton-electron state in the classical Bohr model of the H atom. By using special relativity, the equivalence principle and the de Broglie wavelength equation, we find that when each of the three rotating particles has the same rest mass as the rest mass of a neutrino or an antineutrino ($\sim 0.05\:eV/c^2$) then surprisingly the composite rotating state has the rest mass of the stable baryons, i.e. of the proton and the neutron ($\sim 1 \:GeV/c^2$). This rest mass is due almost exclusively to the kinetic energy of the rotating particles. The results are found to be consistent with the theory of general relativity. The model contains no unknown parameters, describes both asymptotic freedom and confinement and also provides good agreement with QCD regarding the QCD condenstation temperature. Predictions for the thermodynamic and other physical properties of these bound rotational states are compared with experimental values.\\

\end{abstract}

\section{Introduction}
The semiclassical Bohr model for the H atom, first presented a century ago \cite{Bohr1913}, provides quantitative description of all the basic properties of the H atom. In this model one utilizes both the corpuscular and the ondular (wave) nature of the rotating electron. Indeed, by considering the corpuscular nature of an electron of mass $m_e$, Newton's second law for a circular particle motion implies
\begin{equation} 
\label{eq1} 
F=m_e\texttt{v}^2/R.
\end{equation}

Assuming that the force $F$ is described by Coulomb's law, i.e., 
\begin{equation} 
\label{eq2} 
F=e^2/\epsilon R^2,
\end{equation}
eq. (\ref{eq1}) yields
\begin{equation} 
\label{eq3} 
R= e^2/\epsilon m_e\texttt{v}^2.
\end{equation}

The additional equation needed to obtain $\texttt{v}$ and R is obtained by utilizing the ondular nature of the electron, viewed as a standing wave, via the de Broglie wavelength expression
\begin{equation} 
\label{eq4} 
\mathchar'26\mkern-10mu\lambda=R=\frac{n \hbar}{m_e\texttt{v}},
\end{equation}
where $\mathchar'26\mkern-10mu\lambda$ is the reduced de Broglie wavelength (assumed to be equal to the rotational radius $R$), $n$ is a positive integer and $\hbar$ is the reduced Planck constant. 

Upon combining (\ref{eq1})-(\ref{eq4}) one obtains the following well known formulas:
\begin{equation} 
\label{eq5} 
\texttt{v}/c=\frac{e^2}{n\epsilon c\hbar}=\frac{\alpha}{n},
\end{equation}
\begin{equation} 
\label{eq6} 
R=\frac{n^2\hbar}{\alpha m_e c}=n^2a_o,
\end{equation}
\begin{equation} 
\label{eq7} 
\mathcal{H}=-\frac{1}{2n^2}\left[\alpha^2 m_e c^2\right],
\end{equation}
where $\alpha(\approx 1/137.035)$ is the fine structure constant, $a_o$ is the Bohr radius and $\mathcal{H}$ is the Hamiltonian.

A graphical solution of equations (\ref{eq3}) and (\ref{eq4}) is given in Figure 1, which underlines that in the Bohr model both the corpuscular and the ondular nature of the electron are considered, the latter expressed via the de Broglie wavelength equation. This equation played a crucial role in the development of quantum mechanics. 
\begin{figure}[ht]
\begin{center}
\includegraphics[width=0.55\textwidth]{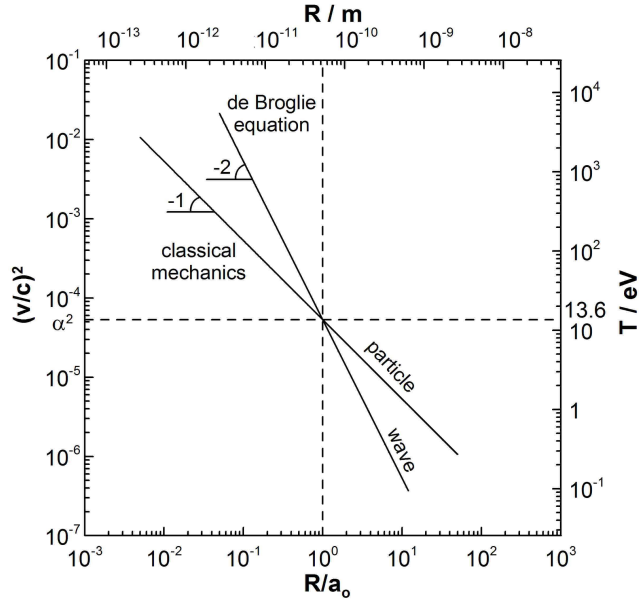}
\caption{Graphical solution of the two equations of the Bohr model, i.e. of the classical mechanical equation $\texttt{v}^2=e^2/\epsilon m_eR$ (eq. \ref{eq3}) and of the de Broglie wavelength equation, for $n=1$, $\texttt{v}^2=\hbar^2/m^2_eR^2$ (eq. \ref{eq4}). The kinetic energy T is computed from $T=(1/2)m_e\texttt{v}^2$, $\texttt{a}_o$ is the Bohr radius $\hbar/m_e\alpha c$.}
\label{fig:1}
\end{center}
\end{figure}

Although the deterministic Bohr model description of the H atom has been gradually replaced by the quantum mechanical Schr\"{o}dinger equation, and is used today mostly for pedagogical purposes, it is worth remembering that the Bohr model (as well as its Bohr-Sommerfeld elliptical orbit extension \cite{Sommerfeld30}), leads to the same level of quantitative description as the Schr\"{o}dinger equation for all the basic properties of the H atom. 
\begin{figure}[ht]
\begin{center}
\includegraphics[width=0.50\textwidth]{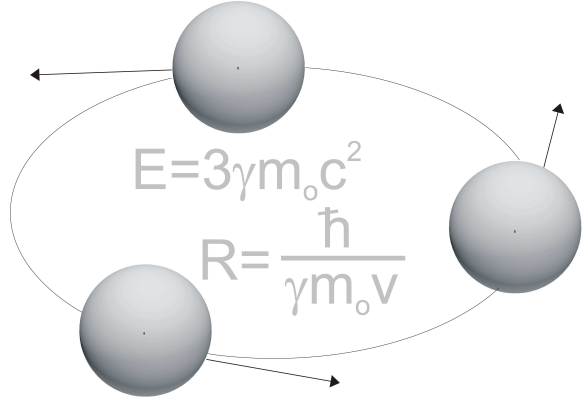}
\caption{Schematic of the model of three rotating neutrinos, together with the expressions for the rest energy and the radius of the bound state.}
\label{fig:2}
\end{center}
\end{figure}

A natural variation of Bohr's model described by equations (\ref{eq3}) and (\ref{eq4}), is to replace the electrostatic attraction by gravity and to examine to what, if any, system such a model may be related to. Thus, as an example, one may consider three light particles (e.g. neutrinos or antineutrinos) each of rest mass $m_o$ (Figure 2) and modify equations (\ref{eq1}) - (\ref{eq4}) to describe the bound rotational state they form via their gravitational attraction. 

We set up to examine what would be the properties of a three-constituent composite particle held together by centripetal gravitational forces using a line of thinking similar to that of the Bohr H atom model.

In the present model for simplicity we consider only the symmetric rotating particle geometry shown in Figures 2 and 3. This geometry, which as shown in section 5.1 is stable, is chosen because it leads to simple analytical expressions which are in good agreement with experiment as shown in sections 4 and 5.  

Also in analogy with the Bohr model, we introduce angular momentum quantization via the use of the de Broglie wavelength equation. More complicated three-body geometries and a full quantum mechanical approach could in principle be tackled via the Faddeev equations which are, however, non-relativistic. \cite{Ahmadzadeh65,Oset12}. In such a case, one should use the gravitational and thus inertial mass of the particles, rather than their rest mass, in generating the attractive potentials, as discussed in section 3.

In order to account for the possibility that in the rotational state the light particles have relativistic velocities  \cite{Torkelsson}, we start from the relativistic equation of motion
\begin{equation} 
\label{eq8} 
\textbf{F}=\gamma m_o\frac{d\textbf{v}}{dt}+\gamma^3m_o\frac{1}{c^2}\left(\textbf{v}  \bullet\frac{d\textbf{v}}{dt}\right)\textbf{v},
\end{equation}
where $\gamma=(1-\texttt{v}^2/c^2)^{-1/2}$ is the Lorentz factor.
\vspace{0.2cm}
\begin{figure}[h]
\begin{center}
\includegraphics[width=0.45\textwidth]{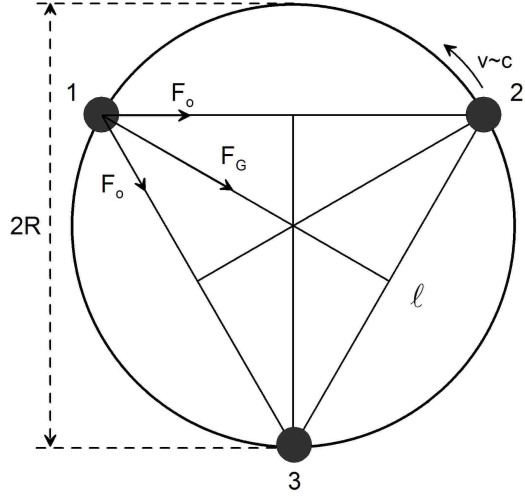}
\end{center}
\caption{Three particles moving at a constant tangential velocity $\textbf{v}$ in a circle of radius $R$ around their center of mass. They are equally spaced. $F_{o}$ is the attraction force between two particles at distance $\ell=\sqrt{3}R$ and $F_{G}$ is the resultant radial force.}
\label{fig:3}
\end{figure}

Since the motion is assumed to be circular, in analogy with (\ref{eq1}) we have 
\begin{equation} 
\label{eq9} 
F=\gamma m_o\frac{\texttt{v}^2}{R}.
\end{equation}

The gravitational centripetal force can be expressed using Newton's universal gravitational law
\begin{equation} 
\label{eq10} 
F=\frac{Gm^2_g}{\sqrt{3}R^2},
\end{equation}
where $m_g$ is the gravitational mass of each particle. The denominator is obtained by noting that the distance $\ell$ between any two of the rotating particles is given by $\ell=\sqrt{3}R$ and that the force is given by $F=2F_{o}cos(\pi/6)=\sqrt{3}F_{o}$, where $F_{o}=Gm^2_o/\ell^2$ is the force exerted between any two of the rotating particles (Figure 3). 

Using the equivalence principle of gravitational and inertial mass, $m_i$, of Eotv\"{o}s and Einstein \cite{Roll1964}, which is known to be valid to 1 part per $10^{12}$ \cite{Roll1964}, i.e. upon using the eq. 
\begin{equation} 
\label{eq11} 
m_i=m_g,
\end{equation}
one obtains in analogy with (\ref{eq3}) the eq.
\begin{equation} 
\label{eq12} 
\gamma m_o\frac{\texttt{v}^2}{R}=\frac{Gm^2_i}{\sqrt{3}R^2}.
\end{equation}

As in the case of the Bohr model for the H atom, the second equation needed to solve for $\gamma$ (thus $\texttt{v}$) and $R$ can be obtained via the use of the de Broglie wavelength equation in the form
\begin{equation} 
\label{eq13} 
R=\mathchar'26\mkern-10mu\lambda=\frac{(2n-1)\hbar}{p}=\frac{(2n-1)\hbar}{\gamma m_o\texttt{v}},
\end{equation}
when $n$ is a positive integer and $p$ is the momentum of each particle. The use of $(2n-1)$ rather than $n$ in this equation can only be justified \textit{a posteriori}, namely in this way better agreement is obtained between certain predictions for the masses of the excited rotational states and experimental values, as presented in section 4. 

Therefore, the only technical problem remaining is order to close the system of equations (\ref{eq12}) and (\ref{eq13}), is to express the inertial mass $m_i$ of the three particles in terms of their rest mass $m_o$ and the particle velocity. 

\section{Inertial and gravitational mass}
The inertial mass $m_i$ of a particle with rest mass $m_o$ is a scalar defined as the ratio of force to acceleration. Thus, when force and acceleration are colinear it is given by the formula
\begin{equation} 
\label{eq14} 
m_i=\frac{\textbf{F}}{d\textbf{v}/dt}=\frac{F}{d\texttt{v}/dt}.
\end{equation}

Recalling that $\textbf{F}$ is the time derivative of the momentum $\textbf{p}$, one obtains
\begin{equation}
\label{eq15}
\textbf{F}=\frac{d\textbf{p}}{dt}=\frac{d}{dt}(m_o\gamma \textbf{v}),
\end{equation}
where $\gamma=(1-\texttt{v}^2/c^2)^{-1/2}$ is the Lorentz factor. One can express the velocity, $\textbf{v}$, in the form
\begin{equation} 
\label{eq16} 
\textbf{v}=\texttt{v}\hat{\textbf{v}},
\end{equation}
where $\hat{\textbf{v}}$ is the unit vector in the direction of $\textbf{v}$. If $\hat{\textbf{v}}$ is fixed, i.e. if
\begin{equation} 
\label{eq17} 
\frac{d\hat{\textbf{v}}}{dt}=0,
\end{equation}
then (\ref{eq15}) becomes
\begin{equation}
\label{eq18}
\textbf{F}=m_o\left[\frac{d}{dt}(\gamma \texttt{v})\right]\hat{\textbf{v}},
\end{equation}
and $\textbf{F}$ and $\textbf{v}$ are colinear.

Using the definition of $\gamma$ one obtains
\begin{eqnarray}
\label{eq19}
\frac{d}{dt}(\gamma \texttt{v})&=&\frac{d\gamma}{dt}\texttt{v}+\gamma\frac{d\texttt{v}}{dt}=\left[\gamma^3\frac{\texttt{v}^2}{c^2}+\gamma\right]\frac{d\texttt{v}}{dt}=\gamma\left[1+\frac{\texttt{v}^2}{c^2}\gamma^2\right]\frac{d\texttt{v}}{dt}\\&=&
\nonumber \gamma^3\frac{d\texttt{v}}{dt}.
\end{eqnarray}

Substituting this expression in (\ref{eq18}) one finds
\begin{equation}
\label{eq20}
\textbf{F}=m_o\gamma^3\frac{d\texttt{v}}{dt}\hat{\textbf{v}},
\end{equation}
thus
\begin{equation}
\label{eq21}
\textbf{F}=m_o\gamma^3\frac{d\textbf{v}}{dt},
\end{equation}
which in view of (\ref{eq14}) yields
\begin{equation}
\label{eq22}
m_i=\gamma^3m_o.
\end{equation}

This result applied to the case of linear motion, is known since Einstein's pioneering special relativity paper \cite{Einstein1905} and the mass $\gamma^3m_o$ is known as the longitudinal mass \cite{Einstein1905,French68,Freund08}.

\section{Arbitrary particle motion and instantaneous inertial reference frames}
We now consider a particle with rest mass $m_o$ and instantaneous velocity $\textbf{v}$ relative to an observer in a reference frame $S$, and we focus on an instantaneous reference frame $S'$ moving with the particle. The instantaneous inertial frame $S'$ is uniquely defined by the instantaneous vector $\textbf{v}$ and thus for this instantaneous frame, $\textbf{v}$ is a constant \cite{French68}. Let $\delta \textbf{v}$ be a small change in the velocity in the same direction with $\textbf{v}$. Then,
\begin{equation}
\label{eq23}
\delta\textbf{v}=\delta\texttt{v}\hat{\textbf{v}}
\end{equation}
and since $\textbf{v}$ is constant,
\begin{equation}
\label{eq24}
\frac{d}{dt}\hat{\textbf{v}}=0.
\end{equation}

Hence, denoting by $\delta \textbf{F}$ the corresponding colinear change in $\textbf{F}$, we find the following:
\begin{eqnarray}
\label{eq25}
\textbf{F}+\delta\textbf{F} &=&\frac{d}{dt}\left[m_o(\gamma + \delta\gamma)(\textbf{v}+\delta\textbf{v})\right]\\&=&
\nonumber  \frac{d}{dt}\left[ m_o\gamma \textbf{v}\right]+\frac{d}{dt} \left[m_o\gamma\delta\textbf{v}+m_o\delta\gamma\textbf{v}\right].
\end{eqnarray}

Thus
\begin{equation}
\label{eq26}
\delta\textbf{F}=m_o\frac{d}{dt}\left[\gamma\delta \textbf{v}+\delta \gamma \textbf{v}\right].
\end{equation}

The definition of $\gamma$ implies the following:
\begin{eqnarray}
\gamma + \delta \gamma &=&\left[1-\frac{(\texttt{v}+\delta \texttt{v})^2}{c^2} \right]^{-1/2}=\left[1-\frac{\texttt{v}^2}{c^2}- \frac{2\texttt{v} \delta \texttt{v}}{c^2} \right]^{-1/2}+\mathcal{O}(\delta \texttt{v})^2=\\&=&
\nonumber \left[\left(1-\frac{\texttt{v}^2}{c^2}\right)\left(1-\frac{2\frac{\texttt{v}}{c^2}\delta \texttt{v}}{1-\frac{\texttt{v}^2}{c^2}}\right)\right]^{-1/2}+\mathcal{O}(\delta \texttt{v})^2=\\&=&
\nonumber \left(1-\frac{\texttt{v}^2}{c^2}\right)^{-1/2}\left[1+\frac{\frac{\texttt{v}}{c^2}\delta \texttt{v}}{1-\frac{\texttt{v}^2}{c^2}}\right]+\mathcal{O}(\delta \texttt{v})^2=\\&=&
\nonumber  \gamma\left[1+\frac{\texttt{v}}{c^2}\gamma^2\delta \texttt{v}\right]+\mathcal{O}(\delta \texttt{v})^2.
\label{eq27}
\end{eqnarray}

Hence
\begin{equation}
\label{eq28}
\delta\gamma=\frac{\texttt{v}}{c^2}\gamma^3 \delta \texttt{v}+\mathcal{O}(\delta \texttt{v})^2.
\end{equation}

Substituting the above in (\ref{eq26}) we find
\begin{eqnarray}
\label{eq29}
\delta\textbf{F} &=& m_o\frac{d}{dt}\left[\gamma \delta \texttt{v}\hat{\textbf{v}}+ \frac{\texttt{v}}{c^2} \gamma^3 \delta \texttt{v}\texttt{v}\hat{\textbf{v}}\right]\\&=&
\nonumber   m_o\frac{d}{dt}\left[\gamma \delta \texttt{v} +\frac{\texttt{v}^2}{c^2}\gamma^3 \delta \texttt{v}\right]\hat{\textbf{v}}.
\end{eqnarray}

But $\texttt{v}$ and $\gamma$ are constants, thus the above equation becomes
\begin{equation}
\label{eq30}
\delta\textbf{F}=m_o\left(\gamma+\frac{\texttt{v}^2}{c^2}\gamma^3\right)\left(\frac{d}{dt}\delta \texttt{v}\right)\hat{\textbf{v}}.
\end{equation}

Consequently
\begin{equation}
\label{eq31}
\delta\textbf{F}=m_o\gamma^3\frac{d}{dt}(\delta \textbf{v}).
\end{equation}
and $\gamma^3m_o$ is again the inertial mass.

Since $\delta \textbf{v}$ and $\delta \textbf{F}$ can be taken to be infitesimally small, it follows that this result, i.e. $m_i=\gamma^3m_o$, is valid for an arbitrary motion of the particle under consideration with an instantaneous velocity $\textbf{v}$.

It is worth noting that both $\delta\textbf{F}$ and $\delta \textbf{v}$ are parallel to $\textbf{v}$. This fact implies that the force $\delta\textbf{F}$ is invariant \cite{French68}, i.e. the same force is perceived in the instantaneous frame $S'$ and in the laboratory frame $S$.

In summary, the longitudinal mass $\gamma^3m_o$ is the inertial mass not only for linear particle motion but also for arbitrary particle motion. Consequently, according to the equivalence principle, $\gamma^3m_o$ is also the gravitational mass for arbitrary motion, including cyclic motion. The latter confirms the assignment of $\gamma^3m_o$ to the gravitationl mass of rotating neutrinos in the recently developed Bohr-type model for the internal structure of hadrons \cite{Vayenas12}. 

We recall that for circular motions, where $\textbf{v}\cdot\frac{d\textbf{v}}{dt}=0$, but also for elliptical and other cyclic motions, one also defines the so called transverse mass $\gamma m_o$ as the ratio of $\textbf{F}$ and $d\textbf{v}/dt$ \cite{French68}. However this ratio, which actually equals the relativistic mass, cannot be used as the inertial mass since in this case force and velocity are not parallel and thus the force is not invariant \cite{French68}. Indeed let $x$ denote the direction of the instantaneous velocity vector $\textbf{v}$ and $y$ denote a direction vertical to it, let $F_x$ and $F_{x'}$ denote the force components in the $x$ direction perceived by a laboratory observer and by the instantaneous frame observer respectively,  and let $F_y$ and $F_{y'}$ denote the correponding force components in the direction $y$. In addition, let $a_x$, $a_{x'}$, $a_y$ and  $a_{y'}$ denote the corresponding accelerations in the $x$ and $y$ directions as perceived by the laboratory and instantaneous frame observers \cite{French68}.

Taking into consideration the well known \cite{French68} equations
\begin{equation}
\label{eq32}
a_x=\frac{1}{\gamma^3}a_{x'}
\end{equation}
and 
\begin{equation}
\label{eq33}
F_x=\gamma^3m_oa_x,
\end{equation}
as well as the equation
\begin{equation}
\label{eq34}
F_{x'}=m_oa_{x'},
\end{equation} 
it follows that 
\begin{equation}
\label{eq35}
F_{x}=\gamma^3 m_o \frac{a_{x'}}{\gamma^3}=m_oa_{x'}=F_{x'}.
\end{equation} 

This striking result shows that the force is invariant in the $x$ direction, i.e. despite the large differences in mass and acceleration, the $x$ component of the force remains the same \cite{French68}.

Similar calculations for the transverse force \cite{French68}, using
\begin{equation}
\label{eq36}
a_{y}=\frac{1}{\gamma^2}a_{y'}, 
\end{equation} 
imply
\begin{equation}
\label{eq37}
F_{y}=\gamma m_o \frac{a_{y'}}{\gamma^2}=\frac{1}{\gamma}m_oa_{y'}.
\end{equation} 
Thus 
\begin{equation}
\label{eq38}
F_{y}=\frac{1}{\gamma}F_{y'},
\end{equation} 
which shows that the force is \textit{not} invariant in the $y$ direction, hence, it cannot be used for computing the inertial mass. On the other hand, eq. (\ref{eq38}) shows that for large $\gamma$, the $F_y$ component vanishes and hence the laboratory observer perceives a nearly linear particle motion and therefore again he observes an inertial mass of $\gamma^3m_o$.

In view of the equivalence principle $(m_i=m_g,$ eq. 11), the consequences of this result, i.e. $m_i=\gamma^3m_o$ (eq. 22) are significant. It implies that the gravitational force between two particles of rest mass $m_o$, velocity $\texttt{v}$ relative to a laboratory observer, and distance $r$ is given by 
\begin{equation}
\label{eq39}
F_G=\frac{Gm^2_o\gamma^6}{r^2}.
\end{equation} 
for arbitrary particle orientation, including the cyclic motion of the present model. 

One might question the appropriateness of using equation (\ref{eq39}), i.e. Newton's universal gravitational law coupled with special relativity (SR), in the present model instead of using general relativity (GR). However as shown in section 5.3, the model results obtained via eq. (\ref{eq39}) can also be obtained via the Schwarzschild geodesics of GR. 

In order to appreciate the magnitude of $F_G$ at highly relativistic velocities it is worth computing the ratio, $\rho$, of this gravitational force between two neutrinos (\ref{eq39}) to the Coulombic force, $F_C$, between two unit charges at the same distance $r$
\begin{equation}
\label{eq40}
\rho=\frac{\epsilon Gm^2_o}{e^2}\gamma^6.
\end{equation}

Using for $m_o$ the most recent Super-Kamiokande value for the heaviest neutrino mass $(0.051\pm0.01\;eV/c^2$ \cite{Mohapatra07}) and setting $\gamma =1$, one finds that $\rho$ is extremely small, i.e. it equals $1.757\cdot 10^{-57}$. However $\rho$ reaches unity for $\gamma =2.88\cdot 10^9$. Using Einstein's equation $E=\gamma m_oc^2$ one computes that the corresponding neutrino energy is $E=146.8\;MeV$ which is in the range of the highest measured neutrino energies in space \cite{Griffiths08}.
\begin{figure}[h]
\begin{center}
\includegraphics[width=0.60\textwidth]{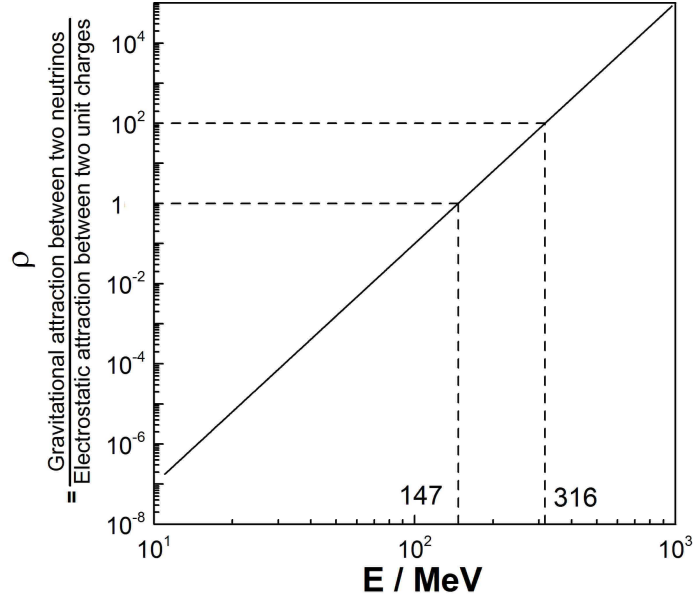}
\end{center}
\caption{Dependence on neutrino energy, $E$ $(=\gamma m_oc^2)$, of the ratio of the gravitational force between two neutrinos of energy $E$ each, to the electrostatic force between two unit charges of opposite sign at the same distance $r$.}
\label{fig:4}
\end{figure}

Therefore one may rewrite equation (\ref{eq40}) using the energy $E$ of the moving particles and Einstein's equation $E=\gamma m_oc^2$ to obtain
\begin{equation}
\label{eq41}
\rho=\frac{\epsilon Gm^2_o}{e^2}(E/m_oc^2)^6=1.757\cdot 10^{-57}(E/m_oc^2)^6.
\end{equation}

Figure 4 provides a plot of eq. (\ref{eq41}), in which we have used again $m_o=0.051\;eV/c^2$ \cite{Mohapatra07} for the neutrino mass. One observes that $\rho$ is negligible for $E<20\;MeV$ but is unity $(\lambda=1)$ for $E=146.8\;MeV$ and becomes two orders of magnitude larger, $\rho =100$, for $E=316\;MeV$. It is interesting to recall that the highest measured neutrino energies in space are in the $100-200\;MeV$ range \cite{Griffiths08} and that the masses of quarks are in the 10-400 $MeV/c^2$ range \cite{Griffiths08}.

\section{Model solution}
Inserting the expression $m_i=\gamma^3m_o$ in eq. (\ref{eq12}) one obtains
\begin{equation}
\label{eq42}
\gamma m_o\frac{\texttt{v}^2}{R}=\frac{Gm_o^2\gamma^6}{\sqrt{3}R^2},
\end{equation}
hence
\begin{equation}
\label{eq43}
R=\frac{Gm_o}{\sqrt{3}c^2}\gamma^5\left(\frac{\gamma^2}{\gamma^2-1}\right).
\end{equation}
\begin{figure}[h]
\begin{center}
\includegraphics[width=0.45\textwidth]{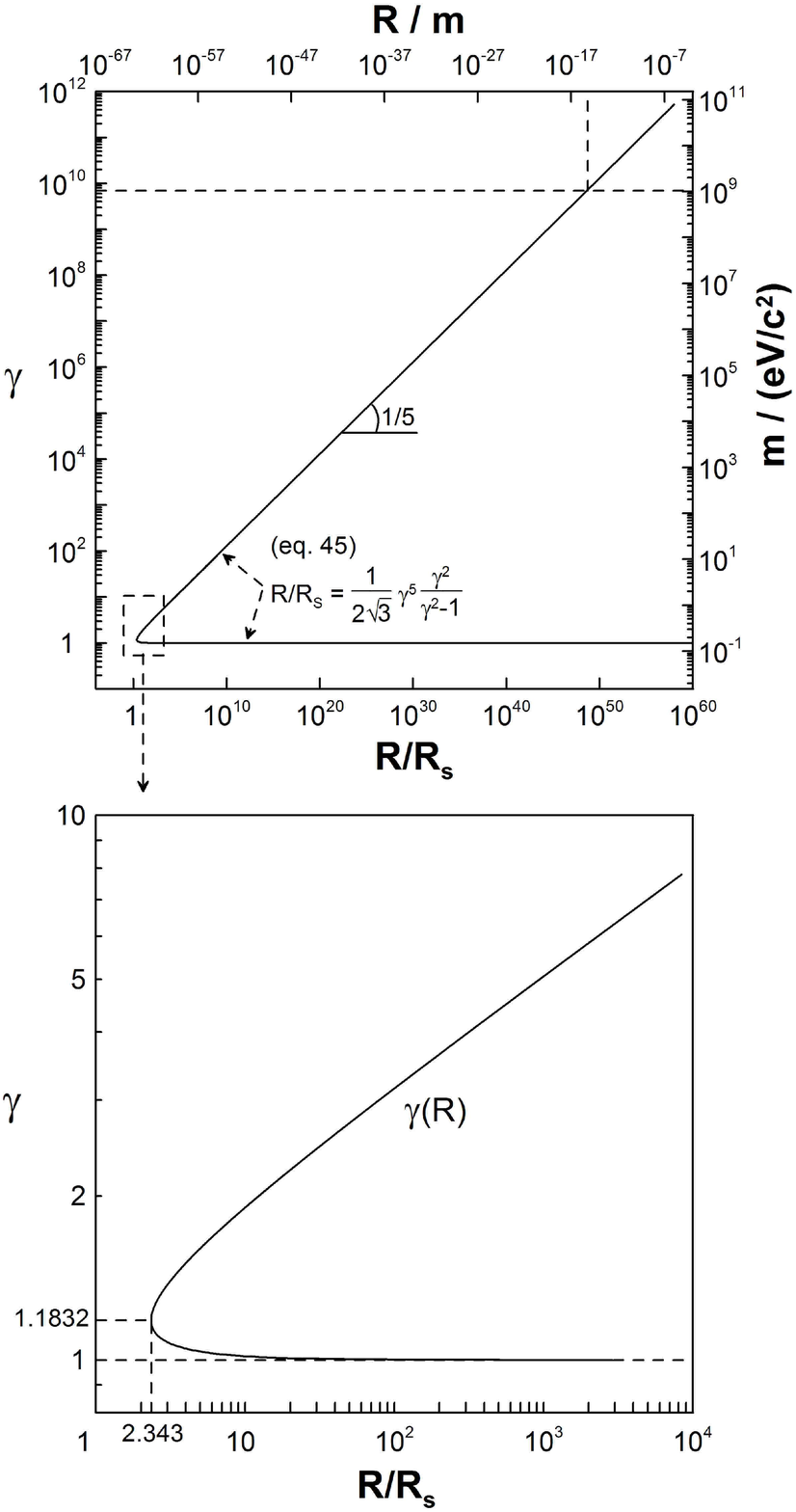}
\end{center}
\caption{Plot of eq. (\ref{eq45}) for R values up to $10^{-5}\;m$ (top), as well for values near $R_{min}$ (bottom). $R_{min}$ denotes the minimum value of $R$. The m axis is constructed from $m=3\gamma m_o$ with $m_o=5.1\cdot 10^{-2}\;eV/c^2$.}
\label{fig:5}
\end{figure}

This equation must be solved in conjunction with eq. (\ref{eq13}), i.e.
\begin{equation}
\label{eq44}
R=\frac{(2n-1)\hbar}{\gamma m_o \texttt{v}}\approx \frac{(2n-1)\hbar}{\gamma m_o c}.
\end{equation}

Using the definition of the Schwarzschild radius, $R_S=2Gm_o/c^2$, one may rewrite equation (\ref{eq40}) in the form
\begin{equation}
\label{eq45}
R=(R_S/(2\sqrt3))\gamma^5\left(\frac{\gamma^2}{\gamma^2-1}\right),
\end{equation}
which reduces to
\begin{equation}
\label{eq46}
R\approx(R_S/(2\sqrt3))\gamma^5,
\end{equation}
for $\gamma>>1$. A plot of equation (\ref{eq45}) is given in Figure 5; solutions for $\gamma$ exist for $R/R_S>2.343$ corresponding to $\gamma=1.1832$. Above this value of $R$, for each $R$ there exist two values for $\gamma$, one $(\gamma\approx 1)$ corresponding to Keplerian orbits ($d\gamma/dR<0$ thus $d\texttt{v}/dR<0$), and one corresponding to relativistic $(\gamma>>1)$ orbits, where, interestingly, $\gamma$ increases with $R$.

The right-side $y$ axis of the top of Figures 5, corresponding to the mass, $m$, of the bound state, is obtained from the equation 
\begin{equation}
\label{eq47}
m=3\gamma m_o,
\end{equation}
which follows from the conservation of energy, i.e. from the equation
\begin{equation}
\label{eq48}
m c^2=3\gamma m_oc^2,
\end{equation}
where $m_o(\approx 0.051\pm 0.01\:eV/c^2)$, is the rest mass of neutrinos \cite{Mohapatra07}. Equation (\ref{eq47}) presents a simple but quite effective mechanism for mass generation by utilizing the kinetic energy of relativistic particles caught in rotational states of larger composite particles.
\begin{figure}
\begin{center}
\includegraphics[width=0.60\textwidth]{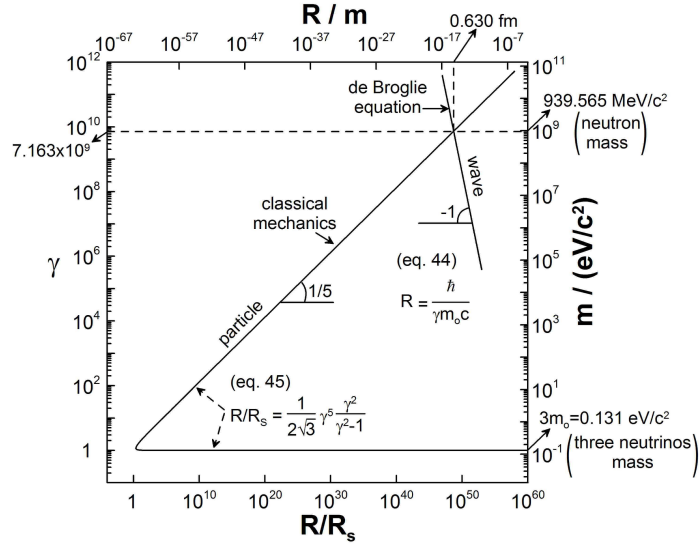}
\caption{Graphical solution of the three neutrino model equations, i.e. of the classical mechanical equation (\ref{eq43}) with its Keplerian $(\gamma<1.1832)$ and non-Keplerian $(\gamma>1.1832)$ branches, and of the de Broglie wavelength equation (\ref{eq44}). The energy or mass axis is obtained from the $\gamma$ axis and from the Einstein equation $E=3\gamma m_oc^2=mc^2$.}
\end{center}
\label{fig:6}
\end{figure}

This equation shows that the rest energy, $mc^2$, of the bound state is the total energy (rest plus kinetic) of the three rotating neutrinos. In fact, since the kinetic energy, $T$, of the three neutrinos is given by
\begin{equation}
\label{eq49}
T=3(\gamma -1)m_oc^2,
\end{equation}
whereas their rest energy is $3m_oc^2$, it follows that for large $\gamma$, as is the case here, the rest energy of the rotational state is overwhelmingly due to the kinetic energy of the rotating neutrinos. As shown in Figure 5, for $R$ values of the order of $1\:fm$, the mass $m$ of the rotational state has already increased from values of the order of $10^{-1}\:eV/c^2$ for $\gamma=1$, to values of the order of $1\:GeV/c^2$. These values lie in the mass range of baryons. 
\begin{figure}[h]
\begin{center}
\includegraphics[width=0.97\textwidth]{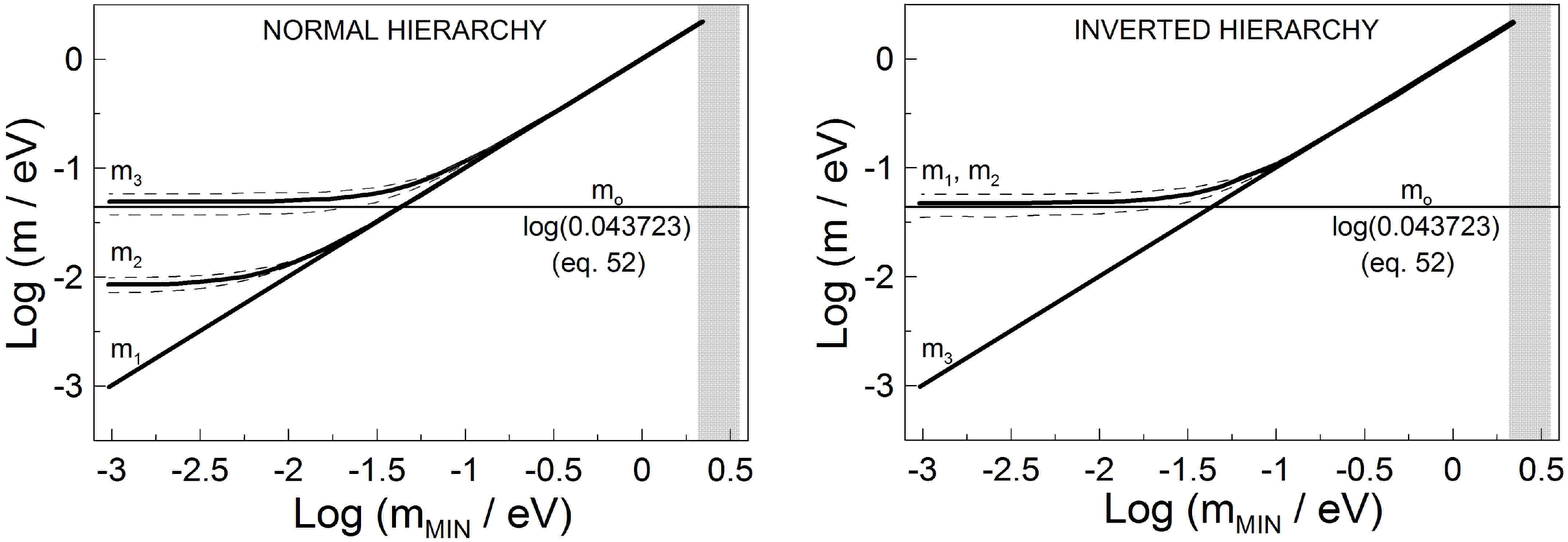}
\caption{The three light neutrino masses as a function of the lightest mass for the normal (left plot) and inverted (right plot) hierarchy, reprinted from ref. \cite{Mohapatra07} and comparison with equation (\ref{eq52}), which implies $m_o=(m_n/3)^{3/2}/(3^{1/8}m^{1/2}_{Pl})=0.043723\;eV/c^2$.}
\label{fig:7}
\end{center}
\end{figure}

The solution of the model is depicted geometrically in Figure 6, which is the same as Figure 5 but also contains the de Broglie equation line of equation (\ref{eq44}) for $n=1$. The intersection of the two lines indeed lies in the $1\:GeV/c^2$ mass region. The exact value of the neutron mass $(m_n=939.565\:MeV/c^2)$ is obtained for a $m_o$ value of $0.043723\:eV/c^2$, which is in good agreement with the experimental value of $0.051\pm 0.01\:eV/c^2$ \cite{Mohapatra07}.

The analytical model solution is found directly by combining equations (\ref{eq44}) and (\ref{eq46}). This yields 
\begin{equation}
\label{eq50}
\gamma=3^{1/12}(2n-1)^{1/6}(m_{Pl}/m_o)^{1/3},
\end{equation}
where $m_{Pl}=(\hbar c/G)^{1/2}$ is the Planck mass. 

The mass $m$ of the rotational state is then obtained from eq. (\ref{eq47}), which yields 
\begin{equation}
\label{eq51}
m=3\gamma m_o=3^{13/12}(2n-1)^{1/6}m_o^{2/3}m_{Pl}^{1/3}.
\end{equation}

Solving (\ref{eq51}) for $m_o$ and setting $n=1$ one obtains 
\begin{equation}
\label{eq52}
m_o=\frac{(m/3)^{3/2}}{3^{1/8}m^{1/2}_{Pl}}.
\end{equation}

\begin{figure}[h]
\begin{center}
\includegraphics[width=0.60\textwidth]{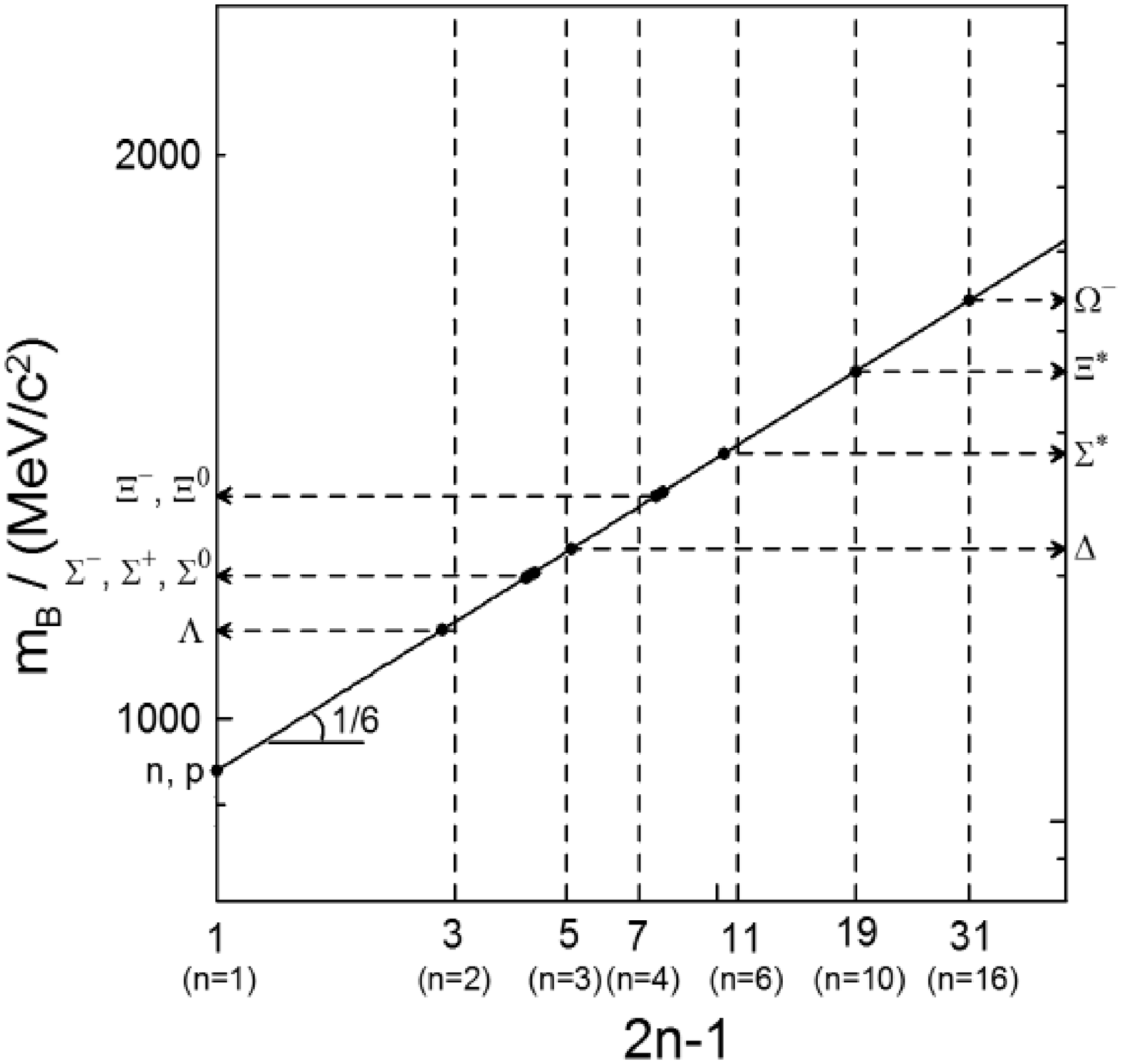}
\caption{Comparison of the masses, $m_B$, of the uncharmed baryons, consisting of u, d and s quarks, \cite{Griffiths08} with equation (\ref{eq53}), i.e. $m_B=m_n(2n-1)^{1/6}$, where $m_n$ is the neutron
mass.} 
\label{fig:8}
\end{center}
\end{figure}

Using $m=m_n=939.565\:MeV/c^2$, which is the neutron mass, one finds $m_o=0.043723\:eV/c^2$, which is in good agreement with the current best estimate of $m_o=0.051(\pm0.01)\;eV/c^2$ for the mass of the heaviest neutrino extracted from the Super-Kamiokande data \cite{Mohapatra07}. This experimental value is computed from the square root of the $\left|\Delta m^2_{23}\right|$ value of $2.6(\pm0.2)\times 10^{-3}(eV/c^2)^2$ extracted from the Super-Kamiokande data for the $\nu_\mu\:\longleftrightarrow\:\nu_\tau$ oscillations \cite{Mohapatra07}.

Actually, as shown in Figure 7 the $m_o$ value of 0.043723 $eV/c^2$ of equation (\ref{eq52}) practically coincides with the currently computed maximum neutrino mass value both for the normal mass hierarchy ($m_3>>m_2>m_1$) and for the inverted hierarchy ($m_1\approx m_2>>m_3$) \cite{Mohapatra07}.

It is worth noting that for any fixed $m_o$ value, equation (\ref{eq51}) can also be written in the form 
\begin{equation} 
\label{eq53}
m=(2n-1)^{1/6}m_n,
\end{equation}
where $m_n$ is the neutron mass. As shown in Figure 8, this expression is also in very good agreement with experiments regarding the masses of baryons consisting of $u$, $d$ and $s$ quarks \cite{Griffiths08}, which follow the $(2n-1)^{1/6}$ dependence of equation (\ref{eq50}) with an accuracy better than 3\% (Fig. 8 and Table 1).

In a similar, and actually simpler, fashion one can formulate the two-rotating particle model \cite{Vayenas12}. The final expression, similar to eq. (\ref{eq51}) for the mass of the confined rotational state is
\begin{equation*} 
m=2\gamma m_o=2^{7/6}(2n-1)^{1/6}m_o^{2/3}m^{1/3}_{Pl}. 
\end{equation*}
For $m_o=0.05(\pm 0.01)\;eV/c^2$ and $n=1$, this gives $m=625(\pm 125)\;MeV/c^2$, which is in the mass range of mesons \cite{Mohr05}.

\begin{table}
\begin{center}
\label{tab:1}
\caption{Experimental \cite{Griffiths08} and computed via eq. (\ref{eq53}) baryon masses}
\begin{tabular}{|p{0.60in}|p{2.2in}|p{1.00in}|p{0.3in}|}\hline
Baryon & Experimental mass value $MeV/c^2$ & $m_n(2n-1)^{1/6}$ & n \\ \hline
\vspace{0.1cm}$N \left\{\begin{array}{l}{p}\\{n}\end{array}\right.$ & \vspace{0.1cm} 938.272\newline 939.565 & \vspace{0.1cm}939.565 &\vspace{0.1cm} 1 \\ \hline 
$\Lambda$ & 1115.68 & 1128.3 & 2 \\ \hline
\vspace{0.1cm}$\Sigma^{+}$\newline $\Sigma^{o}$\newline $\Sigma^{-} $\newline $\Delta$ &\vspace{0.1cm} $\left.\begin{array}{l}1189.37\\1192.64\\1197.45\\1232\end{array}\right\}$\vspace{0.1cm} &\vspace{0.8cm}1228.6 & \vspace{0.6cm}3\\ \hline 
$\Xi ^{o} $\newline $\Xi ^{-} $ & 1314.8\newline 1321.3 & 1299.5 & 4\\ \hline
$\Sigma *$ & 1385 & 1401.2 & 6 \\ \hline
$\Xi *$ & 1533 & 1534.7 & 10 \\ \hline
$\Omega ^{-}$ & 1672 & 1665.3 & 16 \\ \hline
\end{tabular}
\end{center}
\end{table}

\section{Other properties of the rotational states}
The present model implies that the bound rotational state formed by three gravitationally attracting particles, each with rest mass $m_o$ equal to that of a neutrino with $m_o=0.043723\; eV/c^2$, has a rest mass of 939.565 $MeV/c^2$, equal to that of the neutron. This surprising result could be fortuitous. It is therefore useful to examine some other key predictions of the above composite rotational particles and compare them with the corresponding experimental values. 

\subsection{Potential, translational and total energy}
In order to compute the binding energy of the bound state it is necessary to return to the gravitational centripetal force expression eq. (\ref{eq42}) and to use eq. (\ref{eq46}) in order to eliminate $\gamma $ in eq. (\ref{eq42}). In this way, one obtains
\begin{equation} 
\label{eq54} 
F_{G}=m_{o} c^{2} \left(\frac{2\sqrt{3} }{R_{S} } \right)^{1/5} \frac{1}{R^{4/5}}.
\end{equation}

The force equation (\ref{eq54}) refers to circular orbits only, thus it defines a conservative force, since the work done in moving the particles between two points $R_1$ and $R_2$, corresponding to two rotational states with radii $R_1$ and $R_2$, is independent of the path taken. The force vector orientation is also given, pointing to the center of rotation, therefore one can define a conservative vector field as the gradient of a scalar potential, denoted by $V_G(R)$. The latter is the gravitational potential energy of the three rotating particles and corresponds to the energy associated with the transfer of the particles from the minimum circular orbit of radius $R_{\min }$ to an orbit of radius $R$. The function $V_G(R)$ is obtained by integrating eq. (\ref{eq54}):
\begin{eqnarray}
\label{eq55}
V_{G}(R)-V_{G}(R_{min})&=&-\int_{R_{min}}^{R}F_{G}dR'=\\&=&
\nonumber -5m_{o}c^{2} \left(\frac{2\sqrt{3}}{R_{S}}\right)^{1/5}\left(R^{1/5}-R_{\min}^{1/5}\right).
\end{eqnarray}

Noting that $R_{min}=2.343R_S$ (Fig. 5) and that the value of the Schwarzchild radius $R_S$ for neutrinos is extremely small $(\sim 10^{-63}\:m)$, it follows that for any realistic $R$ value (e.g. for a value above the Planck length constant of $10^{-35}\:m)$, equation (\ref{eq55}) reduces to
\begin{equation} 
\label{eq56}
V_{G}(R)=-5m_{o}c^{2}(2\sqrt{3})^{1/5}(R/R_{S})^{1/5}.
\end{equation}

In view of Eq. (\ref{eq46}) one can rewrite equation (\ref{eq56}) in the form
\begin{equation} 
\label{eq57}
V_{G}(R)=-5\gamma m_{o} c^{2}.
\end{equation}

On the other hand, the total kinetic energy $T$ of the three rotating neutrinos is given by eq. (\ref{eq49}).

Thus one may now compute the change in the Hamiltonian $\mathcal{H}$ denoted by $\Delta \mathcal{H}$, i.e. the change in the total energy of the system upon the formation of the rotational bound state of the three originally free neutrinos. The Hamiltonian $\mathcal{H}$ is the sum of the relativistic energy $E=3 \gamma m_oc^2$, and of the potential energy $V_G$. The relativistic energy is the sum of the rest energy $3m_oc^2$, and of the kinetic energy $T$. Denoting by f and i the final and initial states (i.e. the three free non-interacting neutrinos at rest and the bound rotational state) and by (RE) the rest energy, one obtains:
\begin{eqnarray} 
\label{eq58}
\Delta \mathcal{H}&=&\mathcal{H}_{f}-\mathcal{H}_{i}=\\
\nonumber &=&\left[(RE)_{f}+T_{f}+V_{G,f}\right]-\left[(RE)_{i}+T_{i}+V_{G,i} \right]=\\
\nonumber &=&\left[3m_{o}c^{2}+3(\gamma -1)m_{o}c^{2}-5\gamma m_{o}c^{2}\right]-3m_{o} c^{2}=\\ 
\nonumber &=&\Delta T+\Delta V_{G} =-(2\gamma +3)m_{o} c^{2} \approx -2\gamma m_{o} c^{2}
\end{eqnarray}
where the last equality holds for $\gamma \gg 1$.

The negative sign of $\Delta \mathcal{H}$ shows that the formation of the bound rotational state starting from the three initially free neutrinos happens spontaneously, is exoergic $(\Delta \mathcal{H}<0)$, and the binding energy $BE =-\Delta \mathcal{H}$, equals $2\gamma m_oc^2$.

\subsection{Thermodynamic properties}
It follows from (\ref{eq48}) and from (\ref{eq58}) that the binding energy $BE$ is given by
\begin{equation} 
\label{eq59}
BE=-\Delta \mathcal{H}=(2/3)mc^{2}.
\end{equation}

Thus, the binding energy per light particle is $(2/9)mc^2$, which for $m=m_p=938.272\;MeV/c^2$, the proton mass, gives an energy of 208 $MeV$, in good qualitative agreement with the estimated particle energy of 150-200 $MeV$ at the transition temperature of QCD \cite{Braun07}. Furthermore, it gives an even better agreement with the QCD scale of $217 \pm 25\;MeV$ \cite{Wilczek04}.

One may note here that the potential energy expression (\ref{eq57}) can be shown easily not to depend on the number, $N$, of rotating particles. On the other hand, the kinetic energy T is a linear function of $N$, namely $T=N(\gamma -1)m_oc^2$. Thus, it follows from (\ref{eq58}) that stable rotational states $(\Delta \mathcal{H}<0)$ cannot be obtained for $N>5$ since they lead to positive $\Delta \mathcal{H}$. Thus, in addition to the case $N=3$ treated here, the cases $N=2$ and $N=4$ are also interesting cases. For the case $N=2$, in a way similar to that presented in section 4, one finds composite masses, $m$, in the range of mesons, i.e. in the range of $0.5\;GeV/c^2$  \cite{Vayenas12,Griffiths08}.

The change in Helmholz free energy, $F$, can be computed from:
\begin{equation}
\label{eq60}
\Delta F=\Delta \mathcal{H}-\Theta\Delta \mathcal{S}
\end{equation}
where $\Theta$ is the absolute temperature in $K$ and $\Delta \mathcal{S}$ in the entropy change associated with baryon formation from the three neutrinos. The sign of $\Delta \mathcal{S}$ is negative, as three translational degrees of freedom are being lost upon formation of the confined state. Thus to a good approximation it is:
\begin{equation}
\label{eq61}
\Delta \mathcal{S}=-k_bln 3
\end{equation}
where $k_b=1.38\cdot 10^{-23}\;J/K=8.617\cdot 10^{-5}\;eV/K$ is the Boltzmann constant. 

Upon combining eqs. (\ref{eq58}), (\ref{eq60}) and (\ref{eq61}) one obtains that the free energy vanishes, i.e. $\Delta F=0$, at:
\begin{equation}
\label{eq62}
k_b\Theta_{cr}=\frac{(2/3)mc^2}{ln 3}=570.15\;MeV
\end{equation}
where $\Theta_{cr}$ is the critical temperature corresponding to equilibrium between condensed (i.e. confined) and free neutrinos. This temperature is similar to the condensation temperature of the quark-gluon plasma \cite{Braun07,Aoki06,Fodor04} which is estimated to be $1.90\cdot 10^{12}\;K$ \cite{Fodor04}.

It follows from (\ref{eq62}) that the critical kinetic (thermal) energy per particle is:
\begin{equation}
\label{eq63}
T_{cr}=k_b\Theta_{cr}/3=190.05\;MeV
\end{equation}
and consequently the model-computed condensation temperature is
\begin{equation}
\label{eq64}
\Theta_{cr}=2.206\cdot 10^{12}\;K
\end{equation}

The above computed critical or condensation energy and temperature are in good agreement with the predictions of the QCD Theory about the QCD transition energy and temperature (i.e. 160-200 $MeV$ and $\sim 1.9\cdot 10^{12}\;K$ respectively) \cite{Braun07,Aoki06,Fodor04}. Consequently the predictions of the rotating neutrino model are in good agreement with experiment both regarding the QCD scale and the QCD transition energy and temperature. 

\subsection{Confinement and asymptotic freedom}
While the magnitude of the gravitational force acting on the rotating particles increases with decreasing radius $R$ (eq. \ref{eq54}), the absolute value $\left|V_G(R)\right|$ of the gravitational potential energy increases monotonically with increasing $R$ and is unbound (eq. \ref{eq56}). Therefore, equation (\ref{eq56}) describes confinement, which is one of the main characteristics of the strong force \cite{Gross73,Politzer73,Cabibbo75}. The same equation (\ref{eq56}) also describes asymptotic freedom \cite{Gross73,Politzer73,Cabibbo75}, namely the attractive interaction energy becomes very small at short distances, which is a second key characteristic of the strong force \cite{Gross73,Politzer73,Cabibbo75}. Confinement and asymptotic freedom are shown clearly in Figure 9, which depicts the dependence of $\left|V_{G}(R)\right|=-V_{G}(R)$ on $R$. The same figure shows the $R$ dependence of $V_G(R)$, $T(R)$ and $\Delta \mathcal{H}(R)$.
\vspace{0.7cm}
\begin{figure}[ht]
\begin{center}
\includegraphics[width=0.50\textwidth]{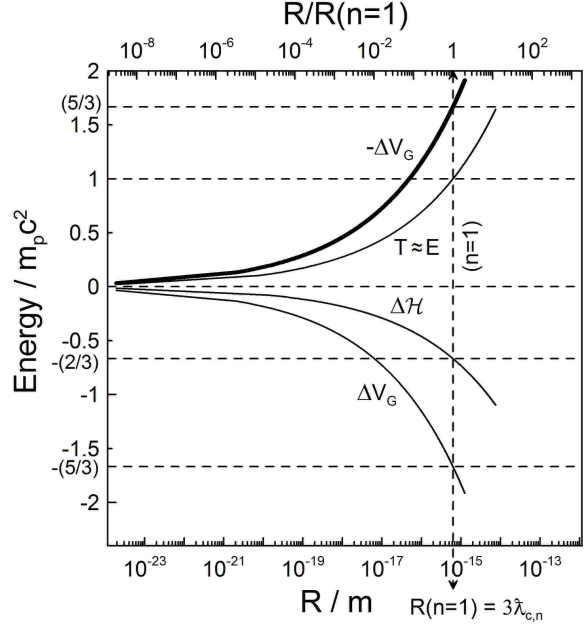}
\caption{Confinement and asymptotic freedom: Plot of eqs \ref{eq49}, \ref{eq56}, \ref{eq58} and \ref{eq44}, the latter for $n=1$, showing the dependence  of the kinetic energy $T$, of the potential energy $\Delta V_G$, and of the Hamiltonian (total energy) $\Delta \mathcal{H}$, on the rotational radius $R$. It is $\Delta \mathcal{H}<0$ for all $R$, indicating that the formation of the bound state is exoergic and thus occurs spontaneously. The total energy $E$ of the bound state equals $3\gamma m_oc^2$ and thus practically coincides with $T$ which equals $3(\gamma -1)m_oc^2$. Note that the $-\Delta V_G$ vs $R$ curve exhibits both asymptotic freedom $(- \Delta V_G \rightarrow 0$ for $R\rightarrow 0)$ and color confinement $(-\Delta V_G \rightarrow \infty$ for $R\rightarrow \infty)$.}
\label{fig:9}
\end{center}
\end{figure}

\subsection{Consistency with general relativity}
It is interesting to examine if the key results of the present model which is based on Newton's gravitational law and special relativity(SR), i.e. equations (\ref{eq39}) and (\ref{eq50}), can be obtained using the theory of general relativity (GR). Thus recalling equations (\ref{eq39}) and (\ref{eq50}), i.e. 
\begin{equation}
\label{eq65}
F_{SR}=\frac{Gm^2_o\gamma^6}{r^2}\quad and \quad \gamma^6_n=3^{1/2}\left(\frac{m_{Pl}}{m_o}\right)^2,
\end{equation} 
it follows that
\begin{equation}
\label{eq66}
\frac{F_{SR}}{F_o}=3^{1/2}\left(\frac{m_{Pl}}{m_o}\right)^2.
\end{equation} 

Assuming for $m_o$ the value $m_o=0.0437\;eV/c^2$, eq. (\ref{eq66}) implies
\begin{equation}
\label{eq67}
\frac{F_{SR}}{F_o}=1.35\cdot 10^{59}.
\end{equation} 

In order to apply the Schwarzschild geodesics equations of GR \cite{Wald84} to the rotating neutrino model, it is first necessary to adjust the physical model of Fig. 3 to the standard geometry of the Schwarzschild metric which involves a light test particle of mass $m_o$ rotating around a central mass $M$. This can be done via the model shown in Figure 10 where the central mass $M$ is fictitious and the value of $M$ is to be specified later. First we note that in the three-rotating particle model the Newtonian, for $\gamma=1$, force exerted to each particle is given by $F=Gm^2_o/(\sqrt{3}R^2)$, see eq. (\ref{eq10}); therefore the Newtonian, for $\gamma=1$, potential energy due to the other two particles is given by
\begin{equation}
\label{eq68}
V_G=-\frac{Gm^2_o}{\sqrt{3}R}.
\end{equation}
\begin{figure}[ht]
\begin{center}
\includegraphics[width=0.45\textwidth]{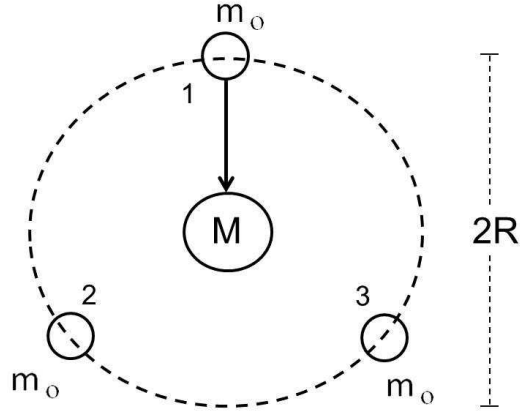}
\caption{GR model for the three-neutrino state.} 
\label{fig:10}
\end{center}
\end{figure}

The same cyclic motion of particle 1 due to particles 2 and 3 (Fig. 10) can be obtained due to the central fictitious mass $M$, provided the value of $G$ is replaced by the value of $G/\sqrt{3}$, since in this case, similarly to eq. (\ref{eq68}), the Newtonian potential energy is given by $V_G=-GM m_o/\sqrt{3} R$. Consequently we consider the one-dimensional Schwarzschild effective potential, $V_{s}(R)$, with $M>>m_{o}$ \cite{Wald84}
\begin{equation} 
\label{eq69}
V_{s}(R)=-\frac{GMm_o}{\sqrt{3}R}+\frac{L^{2}}{2m_{o}R^{2}}-\frac{GML^{2}}{\sqrt{3}c^{2}m_{o}R^{3}},
\end{equation} 
where $L$ is the angular momentum.

Setting $L=\hbar$, we find: 
\begin{equation} 
\label{eq70}
 \frac{V_{s}(R)}{(m_oc^{2}/2)}=\left[-\frac{R_{s,M}}{R} +\frac{a^{2}}{R^{2}}-\frac{a^{2}R_{s,M}}{R^{3}}\right],       
\end{equation} 
where $R_{s,M}$ is the Schwarzschild radius of the central mass $M$ computed with the value $G/\sqrt{3}$, as already noted, and $a$ is the reduced Compton wavelength of the rotating mass $m_o$ \cite{Wald84}, namely $R_{s,M}=2GM/\sqrt{3}c^{2}$ and  $a=\hbar /m_oc$. 

Two circular orbits are obtained when the effective force $dV(r)/dR$ is zero, which upon differentiation of eq (\ref{eq70}) yields
\begin{equation} 
\label{eq71} 
R^\pm=\frac{a^{2}}{R_{s,M}}\left[1\pm \sqrt{1-\frac{3R_{s,M}^{2}}{2a^{2}}}\right].         
\end{equation} 

The smaller one of these roots is unstable (a maximum of $V_s$) and the larger one (a minimum of $V_s$) is very large $(\sim 10^{24}\;m)$ and irrelevant in the present model. In our problem the minimum in $V_s(R)$ occurs on the left boundary, see Figure 11, i.e. at the minimum $R$ value allowed by the Heisenberg uncertainty principle. This may be taken to equal the Compton wavelength, ${\mathchar'26\mkern-10mu\lambda} _{c} $, of the confined mass $M$, where 
\begin{equation} 
\label{eq72} 
\mathchar'26\mkern-10mu\lambda_c=\frac{\hbar}{Mc}.
\end{equation}

We require $R\ge{\mathchar'26\mkern-10mu\lambda}_{c}$, otherwise the uncertainty principle is violated. We also set $M$ to equal the value, $\gamma^3m_o$, of the gravitational mass of each rotating particle found in the SR model. This in conjunction with equation (\ref{eq50}) for $n=1$ gives $M=3^{1/4}m_{Pl}$, thus $M^2=\sqrt{3}m^2_{Pl}$. Therefore
\begin{equation} 
\label{eq73} 
M^2=\sqrt{3}\frac{\hbar c}{G}, \quad  thus \quad \frac{GM}{\sqrt{3}c^2}=\frac{\hbar}{Mc}, \quad i.e. \quad R_{s,M}/2=\mathchar'26\mkern-10mu\lambda_c,
\end{equation} 
as shown in Figure 11.
\begin{figure}[ht]
\begin{center}
\includegraphics[width=0.60\textwidth]{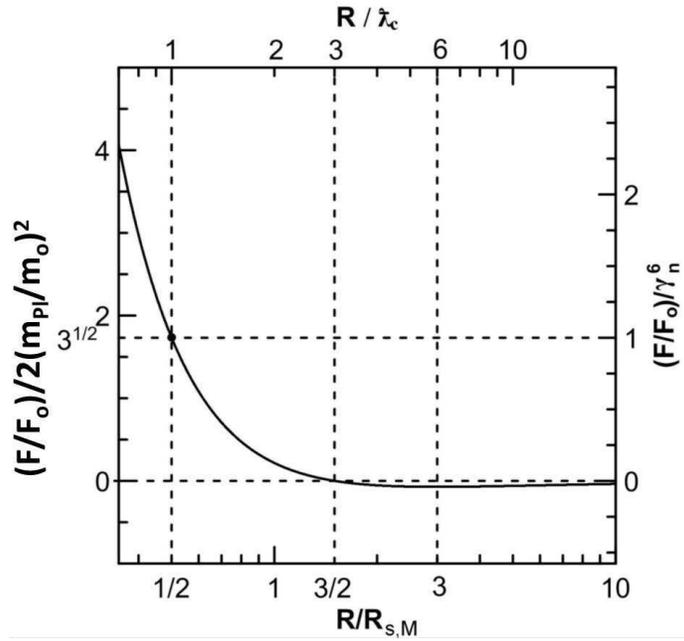}
\caption{Plot of equation (\ref{eq75}) for $M=3^{1/4}m_{Pl}$.} 
\label{fig:11}
\end{center}
\end{figure}

Differentiation of Eq. (\ref{eq70}) allows for the computation of the effective force, $F_s$, acting on each particle:
\begin{equation} 
\label{eq74} 
\frac{F_s}{m_oc^2/2R_{s,M}}=\left(\frac{R_{s,M}}{R}\right)^2+\frac{a^2}{R^2_{s,M}}\left[-2\left(\frac{R_{s,M}}{R}\right)^3+3\left(\frac{R_s}{R}\right)^4\right].
\end{equation} 

Denoting by $F_{GR}$ (for general relativity) the last term in eq. (\ref{eq74}), dividing by the Newtonian force, $F_o$, corresponding to the first term in the same equation, and using the definitions of $a$ and $R_{s,M}$ with $M=3^{1/4}m_{Pl}$, one obtains
\begin{equation} 
\label{eq75} 
\frac{F_{GR}}{F_o}=3^{1/2}\frac{m^2_{Pl}}{4m^2_o}\left[-2\left(\frac{R_{s,M}}{R}\right)+3\left(\frac{R_{s,M}}{R}\right)^2\right].
\end{equation} 

A plot of eq. (\ref{eq75}) is given in Figure 11. For $R=R_{s,M}/2$ this equation gives:
\begin{equation} 
\label{eq76} 
\frac{F_{GR}(R_{s,M} /2)}{F_{o}(R_{s,M}/2)}=2\sqrt{3}\frac{m_{Pl}^{2}}{m_{o}^{2}}.
\end{equation} 
This is in good agreement, differing only by a factor of two, from the value computed from special relativity, namely
\begin{equation}
\label{eq77} 
\frac{F_{SR}}{F_{o}}=\sqrt{3}\frac{m_{Pl}^{2}}{m_{o}^{2}}=\gamma^6_n,
\end{equation} 
obtained via SR, where $\gamma _{n} =m_{n} /3m_{o} $.

The radius $R=\mathchar'26\mkern-10mu\lambda_{c}=(R_{s,M}/2)$ defines a circular rotational state where both GR and quantum mechanics (Compton wavelength equation) are satisfied. At this point there is a net inward force, since $dV_{s}/dR$ is not zero, but this force is counterbalanced by an equal force created by the uncertainty principle via the Compton wavelength requirement. 

\subsection{Radii and Lorentz factors \textbf{$\gamma$}}
The rotational radius $R$ of the bound state computed from equation (\ref{eq44}) for $n=1$, 
\begin{equation} 
\label{eq78}
R(n=1)=\frac{3 \hbar}{m_nc}=0.631\;fm,
\end{equation}
is the neutrino de Broglie wavelength in the bound state and equals three times the neutron Compton wavelength. This value is in a very good agreement with the experimental proton and neutron radii values which lie in the 0.6 - 0.7 $fm$ range. 

For $n>1$, the corresponding $R(n)$ values can be computed from equation (\ref{eq44}), 
\begin{equation} 
\label{eq79}
R=\frac{(2n-1)\hbar}{\gamma m_oc}.
\end{equation}
By accounting for the $\gamma$ dependence on $(2n-1)$ given by equation (\ref{eq50}),
\begin{equation} 
\label{eq80}
\gamma(n)=(2n-1)^{1/6}\gamma(n=1)=7.169\cdot 10^9(2n-1)^{1/6}, 
\end{equation}
one obtains
\begin{equation} 
\label{eq81}
R(n)=(2n-1)^{5/6}R(n=1)=0.631(2n-1)^{5/6}fm.
\end{equation}

The $\gamma(n)$ values correspond to $\sim$ 300 $MeV$ neutrinos. The radii $R(n)$ lie in the range of hadron, e.g. proton or neutron, radii.

\subsection{Lifetimes and rotational periods}
The period of rotation $\tau(n)$ of the rotating particles within the composite state, $2\pi R/\texttt{v}\sim 2\pi R/c$, is given via eq. (\ref{eq81}), by
\begin{equation} 
\label{eq82}
\tau (n)=(2n-1)^{5/6} \tau _{p}=(2n-1)^{5/6} 6.6\cdot 10^{-24} \;s,
\end{equation}
where $\tau_{p}=2\pi R_{p}/c=6.6\times 10^{-24} \;s$ is the rotation period for the proton or the neutron.  The time interval $\tau(n)$ provides a rough lower limit for the lifetime of the composite particles. Indeed, all the known lifetimes of the baryons are not much shorter than this estimate. The lifetime of the $\Delta $ baryons, which is the shortest, is $5.6\cdot 10^{-24} s$ \cite{Griffiths08}.

\subsection{Spins and charges}
Neutrinos are fermions with spin $1/2$ \cite{Griffiths08} and thus one may anticipate spin of $1/2$, or $3/2$ for the composite states formed by three neutrinos. Indeed, most baryons have spin $1/2$ and some have spin $3/2$ \cite{Griffiths08}. If the bound state discussed in the present model involves two neutrinos and one antineutrino, then a spin of $1/2$, that of a neutron, can be anticipated for the bound state.  

Several baryons are charged, such as the proton. Others, such as the neutron, carry no net charge but are known to have an internal charge structure, positive near the center and the external surface, negative in between \cite{Griffiths08}. One can only speculate about the exact location of charge in the rotating neutrino model. It has been discussed \cite{Vayenas12v4,Vayenas12} that positive and negative charge may be generated via the  $\beta$-decay reaction:
\begin{equation} 
\label{eq83}
n\rightarrow p^++e^-+\overline{\nu}_e,
\end{equation}

This is, however, too general and does not provide any physical model about the charge distribution in protons and neutrons. In formulating such a model one should try to account for (a) the actual difference in the masses of reactants and products of (\ref{eq83}), i.e. $m_n-m_p-m_e=0.782\;MeV/c^2$ (b) the values of the magnetic moments of the proton (2.79 $\mu_N$, where $\mu_N=e\hbar/2m_p$ is the nuclear magneton) and of the neutron (-1.913 $\mu_N$). Such a speculative model may involve the following three steps: (1) In the first step a positron $(e^+)$ is trapped in the center of the rotating neutrino ring. It is straightforward to show that the relativistic gravitational force between the positron and the neutrino ring is negligible in comparison to the force keeping the ring neutrinos together (2) in a second step the positron charge is delocalized to generate fractional charges (e/3) on the three electrically polarizable \cite{Mohapatra07} neutrinos which have de Broglie wavelengths extending to the center of the ring. These charges are similar to the assumed charge values of quarks. The resulting structure has a total charge $e$ and a magnetic moment $\mu=(1/2)qR\texttt{v}=(1/2)e(3\hbar/m_pc)c=3\mu_N$ and thus can be reasonably assumed to represent a proton. (3) In a third step a neutron can be generated by capturing an electron and an antineutrino, the latter carrying the energy required for the reverse $\beta$-decay reaction (\ref{eq83}) to occur. The electron charge is added to one of the three preexisting $(e/3)$ charges, so that the resulting charges are $e/3$, $e/3$ and $-2e/3$, identical to the opposite charges of the udd quarks in a neutron. 

Such a charge model is of course highly speculative but can in general account for the non-integer (2/3 and -1/3) charge values of quarks \cite{Griffiths08,Schwarz04} and also allows, via Coulomb's law, for an estimation of the difference in potential energy and thus in the mass between a neutron and a proton $(\sim 1.3\;MeV/c^2)$ \cite{Vayenas12}. Thus considering the Coulombic potential energy, $V_c$, of the electrostatic interaction between three e/3 charges at a distance $\ell =\sqrt{3}R$ for the proton case one computes
\begin{equation} 
\label{eq84}
V_{C,p}=3\frac{(1/9)e^{2}}{\sqrt{3}\varepsilon R}=\frac{\alpha}{9\sqrt{3}}m_nc^2=0.44\;MeV/c^2.
\end{equation}
while for the case of the neutron, which according to the model involves two e/3 charges and one -2e/3 charge, one computes 
\begin{equation} 
\label{eq85}
V_{C,n}=\frac{e^{2}}{\sqrt{3}\varepsilon R}\left[(1/9)-2(2/9)\right]=-\frac{e^{2}}{3\sqrt{3}\varepsilon R}=-\frac{\alpha}{9\sqrt{3}}m_nc^2=-0.44\;MeV/c^2,
\end{equation}

Consequently it is $V_{C,p}-V_{C,n}=0.88\;MeV/c^2$. Thus considering the condition $\Delta\mathcal{H}=0$ for the $\beta$-decay reaction (\ref{eq83}) and neglecting the kinetic energy of the antineutrino produced by the $\beta$-decay one computes
\begin{equation} 
\label{eq86}
m_n-m_p-m_e=(V_{C,p}-V_{C,n})=0.88\; MeV,
\end{equation}
vs 0.782 $MeV$ which is the experimental value. Thus using equation (\ref{eq79}) to estimate the proton mass from $m_n=939.565\;MeV/c^2$ and $m_e=0.511\;MeV/c^2$ one computes $m_p=938.17\;MeV/c^2$ vs 938.272 $MeV/c^2$ which is the experimental value \cite{Mohr05}.

Although this speculative electrostatic model involves some rough approximations, it nevertheless predicts that the neutron mass is larger than the proton mass and that the difference in these masses is of the order of 1 $MeV/c^2$, as experimentally observed \cite{Mohr05}. The model also leads to good estimates of the proton and neutron magnetic dipole moments as described below. 

It is worth noting that, if the Coulomb interaction is taken into consideration, the symmetry of the configuration of Fig. 2 is broken as not all three charges are the same. Although the deviation from three-fold symmetry is small (since the Coulombic energy is small), and thus one may still use with good accuracy eq. (\ref{eq85}) to estimate the attractive interaction between the three particles forming a neutron, it is conceivable that this broken symmetry may be related to the relative instability of the neutron (lifetime 885.7 $s$) vs the proton (estimated lifetime $\sim 10^{32}\;s$ \cite{Griffiths08}).

\subsection{Magnetic moments}
It is interesting to compute the magnetic dipole moments, $\mu$, of the bound rotational states using the charge assumptions of the previous section 5.7. Recalling the definition of $\mu=(1/2)qR$v and assuming positive spin components of all three (anti)neutrinos in the z-axis of the rotating proton state, it follows that
\begin{equation} 
\label{eq87}
\mu_{p}=(1/2)eRc\left[(1/3)+(1/3)+(1/3)\right]=(1/2)eRc .
\end{equation}

Upon substituting $R=R_{p}=0.631 \; fm$, one obtains
\begin{equation}
\label{eq88}
\mu_{p}=15.14\cdot 10^{-27} {\rm \; J/T\; \; \; \; \; (=3\mu_N)},
\end{equation}
where $\mu _{N} $ is the nuclear magneton $(5.05\cdot 10^{-27} {\rm \; }J/T)$. This value differs less than 8\% from the experimental value of $14.10\cdot 10^{-27} {\rm \; J/T\; \; (i.e.\; 2.79\; }\mu _{{\rm N}} )$ \cite{Mohr05}.

In the case of the neutron one may assume negative spin component in the z-axis for the particle with charge $-2e/3$ and opposite spin components of the two $e/3$ charged particles in the z-axis of the rotating proton state to obtain
\begin{equation}
\label{eq89}
\mu_{n}=(1/2)eRc\left[(-2/3)+(1/3)-(1/3)\right]=-(1/3)eRc.
\end{equation}

Upon substitution of $R=0.631\; fm$ one finds
\begin{equation} 
\label{eq90}
\mu_{n}=-10.09\cdot 10^{-27}{\rm \; J/T}=-2\mu_{N},
\end{equation}
which is in very good agreement with the experimental value of $-9.66\cdot 10^{-27} {\rm \; J/T\; }(=-1.913\mu _{N} )$.

This good agreement seems to imply that the spin contribution of the light particles to the magnetic moment of the rotating state is small, and only the sign of the spin projection on the rotational axis (spin up or spin down) is important.

\subsection{Inertial mass and angular momentum}
Interestingly, it follows from equation (\ref{eq47}) that in the case of the neutron or proton $(n=1)$, the inertial and gravitational mass of each rotating particle, $\gamma^3m_o$, is related to the Planck mass, $m_{Pl}=(\hbar c/G)^{1/2}$, via a very simple equation, namely via the eq 
\begin{equation}
\label{eq91}
\gamma^3m_o=3^{1/4}m_{Pl}=3^{1/4}\left(\frac{\hbar c}{G}\right)^{1/2}=1.607\cdot 10^{19}\quad GeV/c^2.
\end{equation}

This provides an interesting direct connection between the Planck mass and the gravitational mass of the rotating neutrino model. The scale of gravity is generally expected to reach that of the strong force at energies approaching the Planck scale $(\sim 10^{19}\;GeV)$ \cite{Schwarz04}, which is in good agreement with the results of the model, see eq. (\ref{eq91}).
\begin{table}[ht]
\centering
\caption{Rest, relativistic, inertial and gravitational mass of the rotating neutrinos for $n=1$.}
\label{tab:2}
\begin{tabular}{p{5.4cm}p{1.8cm}p{3.3cm}p{3.4cm}}
\hline\noalign{\smallskip}
 & Symbol & Value\\
\hline\noalign{\smallskip}
Rest mass & $m_{o}$  & $0.043723\;eV/c^2$\\
Relativistic mass & $\gamma m_{o}$ & $313.188\;MeV/c^2$ & (quark mass range)\\
Inertial mass or gravitational mass & $\gamma^3m_o$ & $1.60692\cdot 10^{19}\;GeV/c^2$ & (Planck mass range)\\ 
Confined state baryon mass $(n=1)$ &  $m=3\gamma m_o$ &  $939.565\;MeV/c^2$ & (neutron mass)\\
\hline\noalign{\smallskip}
\end{tabular}
\end{table}

Thus, while the relativistic mass $3\gamma m_o$ of the bound state formed by the three neutrinos corresponds to $\sim 939\;MeV/c^2$, the inertial and gravitational mass  $\gamma^3 m_o$ of each of them is in the Planck mass range, i.e. $\sim 10^{19}\;GeV/c^2$ (Table 2). 

It is worth recalling at this point Wheeler's concept of geons \cite{Wheeler55,Misner73}, i.e. of electromagnetic waves or neutrinos held together gravitationally, which had been proposed as a classical relativistic model for hadrons \cite{Wheeler55}. In analogy with eq. (\ref{eq91}), the minimum mass of a small geon formed from neutrinos had been estimated \cite{Wheeler55} to lie in the Planck mass range. The behavior of fast neutrinos and quarks in dense matter and gravitational fields has been discussed for years \cite{Itoh70,Braun-Munzinger}. 

It is interesting to note that when using the inertial or gravitational mass $\gamma^3m_o$ in the definition of the Compton wavelength, $\lambda_c$ of the particle $(=h/mc)$, then one obtains the Planck length $(\sim 10^{-35}\;m)$, but when using the mass corresponding to the total energy of the particles, $3\gamma m_o$, then one obtains the proton or neutron Compton wavelength $(\sim 10^{-15}\;m)$, which is close to the actual distance $(\sim fm)$ between the rotating particles.

The model is also qualitatively consistent with another central experimental observation about the strong force, namely that the normalized angular momentum of practically all hadrons and their excited states is roughly bounded by the square of their mass measured in $GeV$ \cite{Vayenas12,Povh06}. Indeed, from eq. (\ref{eq44}) and (\ref{eq53}) one obtains
\begin{equation}
\label{eq92}
(L/\hbar)/(m/GeV)^2=1.13(2n-1)^{2/3},
\end{equation}
which is in reasonable qualitative agreement with experimental values for small integer $n$ values.

\begin{figure}
\begin{center}
\includegraphics[width=0.5\textwidth]{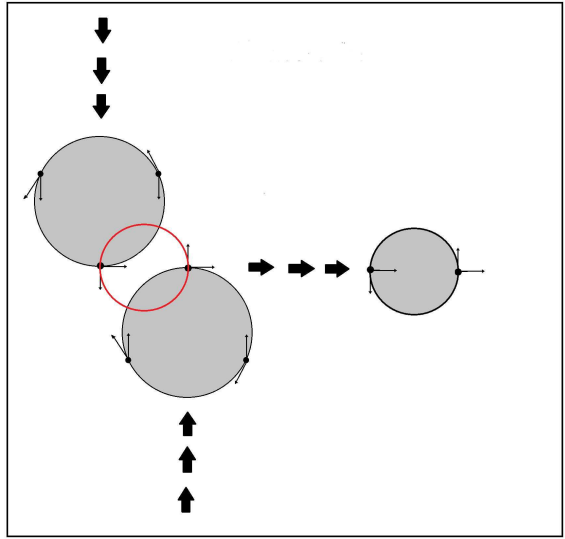}
\end{center}
\caption{Schematic of a possible newly created hadrons jet generation mechanism via the approach of two baryons leading to the formation of a jet in a direction perpendicular to the hadrons motion.}
\label{figure:12}
\end{figure}
\begin{table}[h]
\begin{center}
\caption{Properties of the gravitationally confined three-neutrino states}\par
(using $m_o=0.0437\;eV/c^2$ \cite{Vayenas12} for the neutrino mass)
\begin{tabular}{|p{1.7in}|p{1.4in}|p{1.3in}|} \hline 
\textbf{Property} & \textbf{Value predicted by the model} & \textbf{Experimental value}\\ \hline\hline
\vspace{0.1cm}Neutron rest mass & \vspace{0.1cm} 939.565 $MeV/c^2$ & \vspace{0.1cm}939.565 $MeV/c^2$\\ \hline
\vspace{0.1cm}Baryon binding energy & \vspace{0.1cm} 208 $MeV$ & \vspace{0.1cm} $\sim$150 $MeV$$^*$\\ \hline
\vspace{0.1cm}Reduced de Broglie wavelength\par or radius of ground state & \vspace{0.1cm} 0.631 $fm$ & \vspace{0.1cm} $\sim$0.7 $fm$\\ \hline
\vspace{0.1cm}Minimum lifetime & \vspace{0.1cm} 6.6$\times 10^{-24}$ $s$ & \vspace{0.1cm} 5.6$\times 10^{-24}$ $s$\\ \hline
\vspace{0.1cm}Proton magnetic moment & \vspace{0.1cm} 15.14$\cdot 10^{-27}$ $J/T$ & \vspace{0.1cm}14.10$\cdot 10^{-27}$ $J/T$\\ \hline
Neutron magnetic moment & \vspace{0.1cm} -10.09$\cdot 10^{-27}$ $J/T$ & \vspace{0.1cm} -9.66$\cdot 10^{-27}$ $J/T$\\ \hline
\vspace{0.1cm}Gravitational mass $\gamma^3m_o$ & \vspace{0.1cm} 1.607$\cdot 10^{19}$ $GeV/c^2$ & \vspace{0.1cm} 1.221$\cdot 10^{19}$ $GeV/c^2$\par (Planck mass)\\
\hline
\vspace{0.1cm}Angular momentum & \vspace{0.1cm} 1.13 $(m/GeV)^2$ $\hbar$ $\sim \hbar$ & \vspace{0.1cm} $\sim \hbar$\\ \hline
\end{tabular}
\label{tab:3}
\end{center}
{*}: QCD predicted value at the QCD transition temperature \cite{Braun07}\\ 
\end{table}

\subsection{Scattering cross-sections and hadron jets}
It is possible that the rotating neutrino model may be also able to provide some information regarding the anomalous behavior exhibited by the elastic scattering cross-sections of polarized proton beams, i.e. depending on whether they are parallel or oppositely polarized \cite{Sabbata89}. For example, the cross section for parallel beams, i.e. polarized in the same direction, are up to a factor of four larger than that observed with oppositely polarized beams in the 10 $GeV$ scale \cite{Sabbata89}. This behaviour has been attributed to spin-torsion interactions in the context of supergravity models \cite{Sabbata89}. 

The rotating neutrino model may also provide a qualitative  scheme to account for the emission of jets of newly created hadrons when highly energetic hadrons are forced to collide with each other, such as in the LHC experiments \cite{Khachatryan10}. These jets appear in a direction vertical to the direction of the colliding protons. This behavior could be rationalized as follows:  If two rotating neutrinos, one in each colliding baryon, come close to each other then the gravitational attraction between them can become very large due to their high rotational velocity and small distance, $r$, and in this way the two neutrinos may escape together in a direction vertical to the direction of the colliding baryons (Fig. 12) as experimentally observed.

\subsection{Summary}
In summary, as shown in Table 3, there is good agreement between model and experimental results regarding masses, binding energies, minimum lifetimes, angular momenta and magnetic moments. Furthermore, as already discussed in section 5.3 and Figure 9, the model describes both confinement and asymptotic freedom. There is also good agreement with some key results of QCD regarding the values of the QCD transition temperature and the QCD scale, as discussed in section 5.2. However, QCD can provide a much more complete and exact description of several properties of hadrons than the present simple model.
\begin{table}[h]
\begin{center}
\caption{Comparison of the Bohr model for the H atom and of the Bohr type model for the neutron}
\label{tab:4}
\vspace{\baselineskip}
\includegraphics[width=0.80\textwidth]{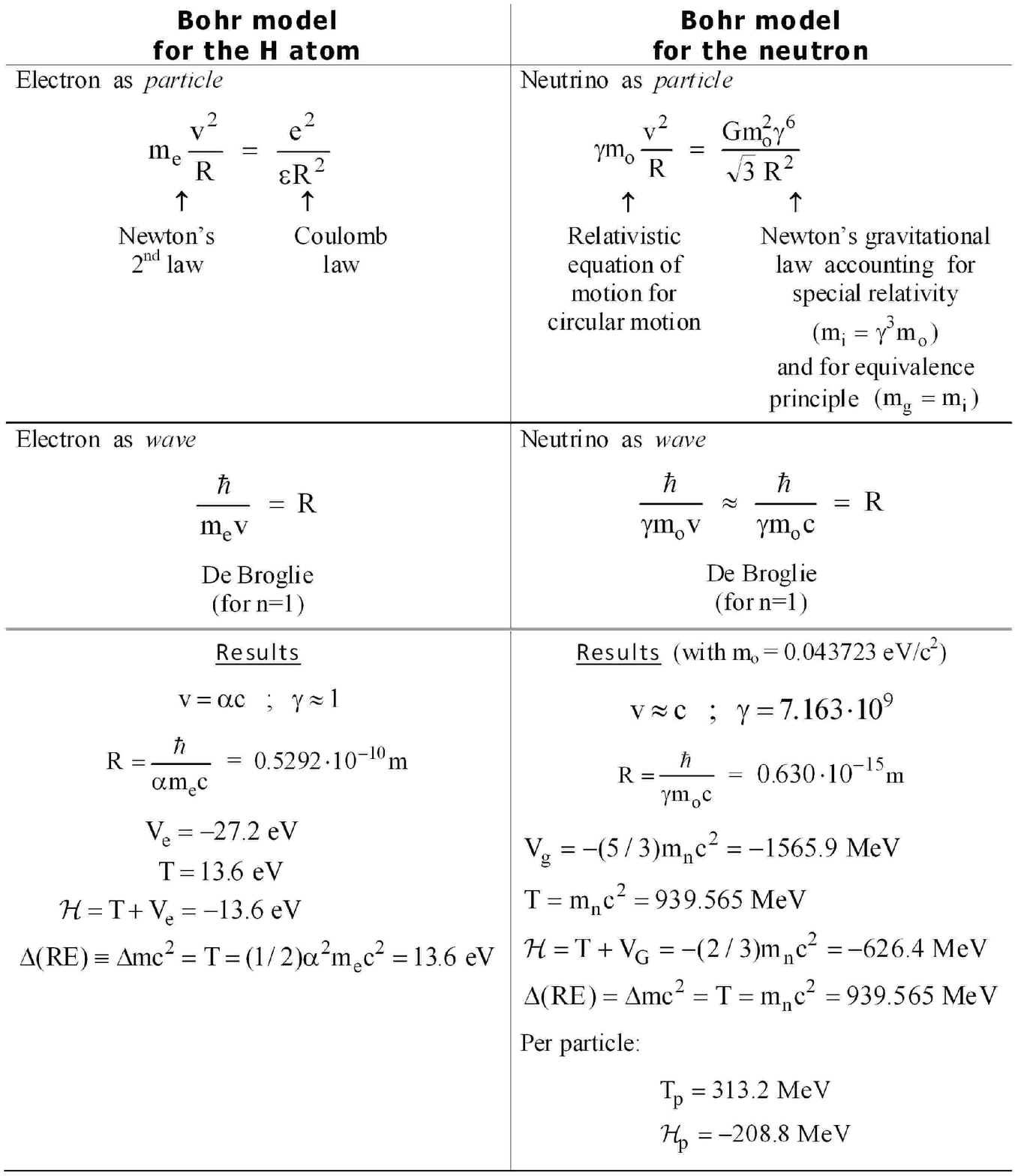}
\end{center}
\end{table}

\section{Conclusions}
A deterministic Bohr-type model can be formulated for the rotational motion of three fast neutrinos using gravity as the attractive force. By employing special relativity, the weak equivalence principle, Newton's gravitational law, and the de Broglie wavelength of the rotating neutrinos which leads to quantization of angular momentum, one finds that the emerging rotational states have, surprisingly, the thermodynamic and other physical properties of baryons, including masses, binding energies, radii, reduced Compton wavelengths, magnetic moments and angular momenta. The key results are also shown to be consistent with the Schwarzschild geodesics of general relativity. Furthermore the model which can be viewed as a simple variation of the Bohr model for the H atom (Table 4), and which contains no unknown parameters, describes both asymptotic freedom and confinement and provides good agreement with QCD regarding the QCD transition temperature and scale.

Since neutrinos and antineutrinos come in three flavors with different masses, it appears worthwhile  to test in the future the usefulness of such deterministic Bohr-de Broglie-type models using various neutrino and antineutrino combinations with gravity as the attractive force, for the possible description of the formation of other composite particles. 

\subsection*{Acknowledgment}
We thank Stefanos Aretakis, Elias Vagenas and Dimitrios Grigoriou for helpful discussions.

\end{document}